\documentclass[12pt]{article}
\usepackage[cp1251]{inputenc}
\usepackage[russian]{babel}
\usepackage{graphicx}

\def\baselinestretch{1.5}
\large \baselineskip = 20 true pt
\begin{document}

\begin{center}
\Large{Influence the effect of long-range interaction on critical
behavior of the three-dimentional systems.}
\end{center}

\begin{center}
{\bf S.V.Belim\\
Omsk State University\\
belim@univer.omsk.su}
\end{center}

\vskip 2mm \begin{center} \parbox{142mm} {\small It Is realized
the theoretic - field description of Ising systems behaviour with
effect of long-range interaction in two-loop approximation in
three-dimensional space with using Pade–Borel resummation
technique. The renorm-group equations are analysed and it is
chosen fixed points, defining critical behaviour of the system. It
Is shown that the influence effect of long-range interaction can
bring as to change the mode of the critical behaviour, so and to
change the type of the phase transition.}
\end{center} \bigskip

Influence effect of long-range interaction, described on greater
distances by sedate law $1/r^{-D-a}$ was explored analytically
within the framework of $\varepsilon$-expansion approach
\cite{1,2,3} and numerically method Monte-Carlo \cite{4,5,6} for
one-dimensional and two-dimensional systems, and shown importance
influences effect of long-range interaction on critical behaviour
of Ising systems for value of the parameter $a<2$. However before
this times are absent article, in which given problem solved
analytically at dimensionality space directly $D=3$. However such
description necessary on the strength of bad convergence of the
rows, got within the framework of $\varepsilon$-expansion. In this
article is conducted description of the critical behaviour
three-dimensional Ising systems with effect of long-range
interaction under different values of the parameter $a$.

Gamilitonian systems with provision for effect of long-range
interaction:
\begin{equation}\label{gam1}
   H=\int d^Dq\Big\{\frac{1}{2}(\tau_0+q^a)\varphi^2+u_0\varphi^4\Big\},
\end{equation}
here $\varphi$ -- fluctuations of the parameter of the order, $D$
-- dimension of space, $\tau_0\sim|T-T_c|$, $T_c$ -- critical
temperature, $u_0$ -- positive constant. Critical behaviour
greatly depends on parameter $a$, assigning velocity of the
decrease the interaction with distance. As it is shown in work
\cite{1} influence effect of long-range interaction greatly under
$0<a<2$, and for $a\geq2$ critical behaviour equivalent of
short-range interaction system. So hereinafter we limit event
$0<a<2$.

Conducting standard renorm-groups procedure on base of the
technology Feynmans diagrams \cite{7} with
$G(\vec{k})=\frac{1}{\tau+|\vec{k}|^a}$, we get expressions for
function $\beta$,$\gamma_\varphi$ и $\gamma_t$, assigning the
differential renorm-group equation.
\begin{eqnarray}
    \beta&=&-(4-D)\Big[1-36uJ_0+1728\Big(2J_1-J_0^2-\frac29G\Big)u^2)\Big],\nonumber\\
    \gamma_t&=&(4-D)\Big[-12uJ_0+288\Big(2J_1-J_0^2-\frac13G\Big)u^2\Big], \nonumber\\
    \gamma_\varphi&=&(4-D)96Gu^2,\\
    J_1&=&\int \frac{d^Dq
    d^Dp}{(1+|\vec{q}|^a)^2(1+|\vec{p}|^a)(1+|q^2+p^2+2\vec{p}\vec{q}|^{a/2})},\nonumber\\
    J_0&=&\int \frac{d^Dq}{(1+|\vec{q}|^a)^2},\nonumber\\
    G&=&-\frac{\partial}{\partial |\vec{k}|^a}\int \frac{d^Dq
    d^Dp}{(1+|q^2+k^2+2\vec{k}\vec{q}|^a)(1+|\vec{p}|^a)(1+|q^2+p^2+2\vec{p}\vec{q}|^{a/2})}\nonumber
\end{eqnarray}
We shall redefy the efficient consn=tant of the interaction:
\begin{equation}\label{vertex}
    v=\frac{u}{J_0}.
\end{equation}
As a result we come to following expression for function $\beta$,
$\gamma_\varphi$ and $\gamma_t$:
\begin{eqnarray}\label{beta}
    \beta&=&-(4-D)\Big[1-36v+1728\Big(2\widetilde{J_1}-1-\frac29\widetilde{G}\Big)v^2)\Big],\nonumber\\
    \gamma_t&=&(4-D)\Big[-12v+288\Big(2\widetilde{J_1}-1-\frac13\widetilde{G}\Big)v^2\Big], \\
    \gamma_\varphi&=&(4-D)96\widetilde{G}v^2,\nonumber\\
    \widetilde{J_1}&=&\frac{J_1}{J_0^2}\ \ \ \
    \widetilde{G}=\frac{G}{J_0^2}.\nonumber
\end{eqnarray}
Such redefining have sense at values $a\leq D/2$. Herewith $J_0$,
$J_1$, $G$ become the dispersing function. Enterring parameter
cutting $\Lambda$ and considering limit of the relations
\begin{eqnarray}\label{lim}
    \frac{J_1}{J_0^2}&=&\frac{\int_0^\Lambda\int_0^\Lambda d^Dq
    d^Dp/((1+|\vec{q}|^a)^2(1+|\vec{p}|^a)(1+|q^2+p^2+2\vec{p}\vec{q}|^a))}
    {\Big[\int_0^\Lambda d^Dq/(1+|\vec{q}|^a)^2\Big]^2},\\
    \frac{G}{J_0^2}&=&\frac{-\partial/(\partial |\vec{k}|^a)\int_0^\Lambda\int_0^\Lambda d^Dq
    d^Dp/((1+|q^2+k^2+2\vec{k}\vec{q}|^a)(1+|\vec{p}|^a)(1+|q^2+p^2+2\vec{p}\vec{q}|^a))}
    {\Big[\int_0^\Lambda d^Dq/(1+|\vec{q}|^a)^2\Big]^2}.\nonumber
\end{eqnarray}
for $\Lambda\rightarrow\infty$ we get the finit expressions.

Values of integrals was found numerically. For $a\leq D/2$
sequence of values was built $J_1/J_0^2$ и $G/J_0^2$ for different
values of $\Lambda$ and was aproximated on infinity.

The mode of the critical behaviour is completely defined fixed
points of renorm-group transformations, which can be found from
condition equality zero $\beta$-function:
\begin{equation}\label{nep}
    \beta(v^*)=0.
\end{equation}
The condition to stability is positive value of derived
$\beta$-functions in fixed point:
\begin{equation}\label{proiz}
   \lambda=\frac{\partial\beta(v^*)}{\partial v}>0.
\end{equation}

Critical exponent $\nu$, the characterizing growing of the radius
of correlation in vicinities of the critical point
$(R_c\sim|T-T_c|^{-\nu})$ is found as:
\begin{eqnarray}
  \nu=\frac12(1+\gamma_t)^{-1}.\nonumber
\end{eqnarray}

Critical exponent $\eta$, the describing behaviour correlation
functions in vicinities of the critical point in space wave vector
$(G\sim k^{2+\eta})$, is defined on base scaling functions
$\gamma_\varphi$: $ \eta=\gamma_\varphi$. Over critical exponents
can be determined from $ \eta$ and $\nu$.

Known that expantion of theories of the indignations are
asymptotic, but constants of the interaction of fluctuation
parameter of the order is enough great to possible was directly
use expressions (\ref{beta}).

So for the reason extractions from the got expressions to
necessary physical information was an aplying method of the using
Pade–Borel resummation technique. Herewith direct and reconversion
Borels transformation are of the form:
\begin{eqnarray}
&& f(v)=\sum\limits_{i} c_{i}v^{i}=\int\limits_{0}^{\infty}e^{-t}F(vt)dt,  \\
&& F(v)=\sum\limits_{i}\frac{\displaystyle c_{i}}{\displaystyle
i!}v^{i}.
\end{eqnarray}
For calculations $\beta$-functions were used аpproximant Pade
[2/1], for $\gamma_t$- and $\gamma_\varphi$-function -- [1/1].

Firm fixed points renorm-groups transformations, derived $\beta$-
functions in fixed to point and critical exponents for values of
the parameter $1.5\leq a \leq 1.9$ are provided in table. For
values of the parameter $0<a<1.5$ exists only Gauss fixed point
$v^*=0$, not being firm. The givenned result will with predictions
$\varepsilon$-expantion \cite{1,2,3}. The absence firm fixed
points is indicative of change the phase turning the second sort
by phase turning the first type.
\begin{table}
\begin{center}
\begin{tabular}{|c|c|c|c|c|c|c|c|c|} \hline
 $a$   & $v^{*}$  &$\lambda$  &$\nu$     &$\alpha$  &$\eta$    &$\gamma$  \\
\hline
 1.5 & 0.015151 &  0.918690 & 0.555566 & 0.333302 & 0.002647 & 1.109661 \\
 1.6 & 0.015974 &  0.874129 & 0.557889 & 0.326333 & 0.003936 & 1.113582 \\
 1.7 & 0.020485 &  0.699732 & 0.567334 & 0.297998 & 0.004862 & 1.131910 \\
 1.8 & 0.023230 &  0.628209 & 0.572714 & 0.281858 & 0.007461 & 1.141155 \\
 1.9 & 0.042067 &  0.683927 & 0.620054 & 0.139838 & 0.013420 & 1.231787 \\
 \hline
\end{tabular} \end{center} \end{table}

The analysis critical exponents shows that at reduction of the
parameter $a$ index $\nu$ decreases that is to say velocity of the
growing of the radius correlation is slowed when approximation to
critical point.

Thereby for three-dimensional Ising systems growing effect
long-range interaction brings first to deceleration of the
velocities of the growing of the radius корелляции in critical
area, but then at importances of the parameter $a<3/2$ to change
the sort of the phase transition.

\def\baselinestretch{1.0}


\begin{thebibliography}{99}
\bibitem{1}
{\it M. E. Fisher, S.-k. Ma and B. G. Nickel}, Phys. Rev. Lett.
{\bf 29} , 917 (1972).
\bibitem{2}
{\it J. Honkonen}, J. Phys. A {\bf 23} , 825 (1990).
\bibitem{3}
{\it E. Luijten, H. Mebingfeld}, Phys. Rev. Lett. {\bf 86} , 5305
(2001).
\bibitem{4}
{\it E. Bayong, H. T. Diep}. Phys. Rev. B. {\bf 9}, 18, 11920
(1999);
\bibitem{5}
{\it E. Luijten}, Phys. Rev. E {\bf 60}, 7558 (1999).
\bibitem{6}
{\it E. Luijten and H. W. J. Blo\"{o}te}, Phys. Rev. B {\bf 56},
8945 (1997).
\bibitem{7}
{\it J.~Zinn-Justin}. Quantum field theory and critical phenomena,
Clarendon Press, Oxford (1989).
\end{thebibliography}
\end{document}